\documentclass[12pt] {article}
\usepackage{amsmath, epsfig, tabularx}

\title{``Stiff'' Field Theory of Interest Rates and Psychological Future Time}
\author{Belal E. Baaquie,\\ Department of Physics, \\
  National University of Singapore; phybeb@nus.edu.sg 
\\Jean-Philippe Bouchaud, \\
Science \& Finance/Capital Fund Management,\\  6-8 Bvd Haussmann, 75009 Paris} 
\begin{document}
\maketitle
\begin{abstract}
The simplest field theory description of the multivariate statistics 
of forward rate variations over time and maturities, involves a quadratic action containing a gradient squared 
rigidity term. However, this choice leads to a spurious kink (infinite curvature) of the normalized correlation function 
for coinciding maturities. Motivated by empirical results, we consider an extended action that contains a squared 
Laplacian term, which describes the bending stiffness of the FRC. With the extra ingredient of a 
`psychological' future time, describing how the perceived time between events depends on the time in the future,
our theory accounts extremely well for the phenomenology of interest rate dynamics. 
\end{abstract}

\maketitle

Forward interest rate $f(t,x)$ is the interest rate, agreed upon at time $t$, 
for an instantaneous loan 
to be taken at future time $x>t$, between $x$ and $x+dx$. At any instant of time $t$, 
the forward rate curve (FRC)
$f(t,x)$ defines a kind of string which moves and deforms with time. Modelling the motion of this curve is of
paramount importance for many financial applications \cite{Hull}: 
pricing of interest rate derivatives, such as `caps' that
limit the rate of loans that individuals take on their housing, interest swaps, risk management 
(asset liability management), etc. The industry standard is the so-called HJM model 
\cite{HJM,Books}. This model 
has recently been generalized in different directions \cite{Ken, String, Gold}, in particular by one of us 
\cite{bebcup,Baaquie,Baaquiesv}, who has proposed a two-dimensional Euclidean quantum field theory for modelling 
the forward interest rate curve. The forward interest rate dynamics has a drift velocity $\alpha(t,x)$ 
and volatility $\sigma(t,x)$; it is convenient to define a driftless noise field $A(t,x)$ by 
\begin{equation}
\label{dfA}
\frac{\partial f}{\partial t}(t,x)=\alpha(t,x)+\sigma(t,x)A(t,x).
\end{equation}
The noise field describes the external shocks of the economy on the 
different maturities of the forward rates; its statistics is 
governed by the exponential of an `action' $S[A]$, which gives the weight of a given path of $A$ in the 
two dimensional space $x,t$, and is defined on the semi-infinite domain $x\ge t$. 
The simplest action that factors in the one dimensional nature of the 
forward rate string was proposed in \cite{bebcup,Baaquie,Baaquiesv}, and reads
\begin{equation}
\label{sa}
S[A]=-\frac{1}{2}\int_{t_0}^{\infty}dt\int_t^{\infty}
dx\left\{A^2(t,x)+\frac{1}{\mu^2}\left(\frac{\partial A(t,x)}{\partial
x}\right)^2 \right\},
\end{equation}
where $\mu$ is a `rigidity' parameter, that gives an elasticity to the 
FRC. 
To eliminate boundary terms from the action we choose to impose 
Neumann boundary condition, i.e.
\begin{eqnarray}
\label{bca}
\frac{\partial A(t,x)}{\partial x}\Big{|}_{x=t}=0,
\end{eqnarray}
corresponding to a parallel motion of the FRC for short maturities, which is reasonable since the spot rate 
$f(t,t)$ is fixed by the central bank, and very short maturities carry no extra risk.

The quantum field theory \cite{zj} needed to describe the statistics of the FRC 
is defined by a functional integral over all variables $A(t,x)$ and 
yields the partition function $Z=\int DA e^{S[A]}$.
The propagator (or noise correlator) is given by 
\begin{equation}
\label{aa}
\langle A(t,x)A(t',x') \rangle =\frac{1}{Z}\int DA~ A(t,x)A(t',x')e^{S[A]}\equiv
\delta(t-t')D(x,x';t).
\end{equation}
Since the above action is Gaussian, the market defined by the above model is complete in the sense that all 
contingent claims can 
be perfectly replicated by hedging appropriately, as in the standard Black-Scholes model. For consistency 
of the description, one should further impose a `martingale' condition that reads \cite{bebcup}  
\begin{eqnarray}
\alpha(t,\theta)=\sigma(t,\theta)\int_{0}^{\theta}d\theta' D(\theta,\theta')\sigma(t,\theta')\nonumber.
\end{eqnarray} 
Note however that this term is usually numerically very small \cite{data}.
With the above choice of the action, the propagator, 
in new co-ordinates $\theta_{\pm}$
is given by
\begin{eqnarray}
\label{lart}
D(\theta_+,\theta_-)=\frac{\mu}{2} [e^{-\mu\theta_+}+e^{-\mu|\theta_-|}]\nonumber\\
\label{thpm}
\end{eqnarray}
with $\theta_{\pm}=\theta \pm \theta'$, $\theta=x-t$  and $\theta'=x'-t$ with $x,x'>t$. 

\begin{figure}[h]
  \centering
  \epsfig{file=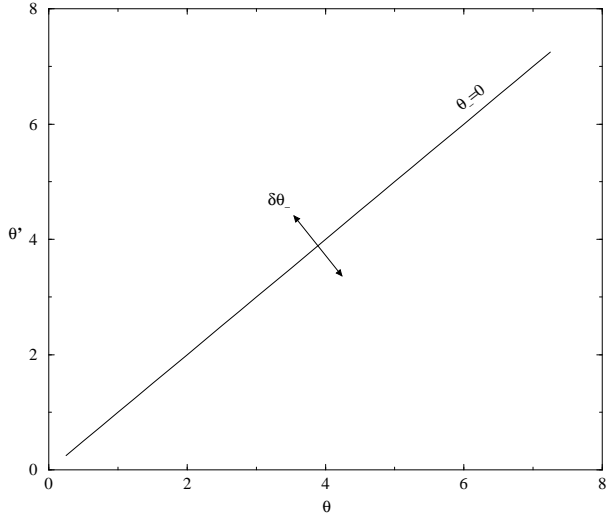, width=7cm, angle=-90}
  \caption{For $\theta_-=\theta-\theta'$, the figure shows that the diagonal axis is given by $\theta_-=0$. 
  The direction of change in $\theta_-$ for constant $\theta_+$, namely $\delta \theta_-$, 
  is orthogonal to the diagonal, as shown in the Figure.}
 \label{thetaminus}
\end{figure}

Note that the slope of the propagator perpendicular to 
$\theta_{-}=0$, as in Figure \ref{thetaminus}, 
is {\it discontinuous} across the diagonal
\begin{equation}
m=\frac{\partial D(\theta_{+};\theta_{-})}{\partial \theta_{-}}\Big{|}_{\theta_{-}=0}
=\frac{\mu^2}{2} \left\{ \begin{array}{ll}
             -1& \theta_-> 0 \\
            +1& \theta_-<0
                \end{array}
\right.
\end{equation} 

All the variants of the propagator based on a gradient squared rigidity term in the action
show a similar infinite curvature singularity along the diagonal \cite{BaaqSri}.

\begin{figure}[h]
  \centering
  \epsfig{file=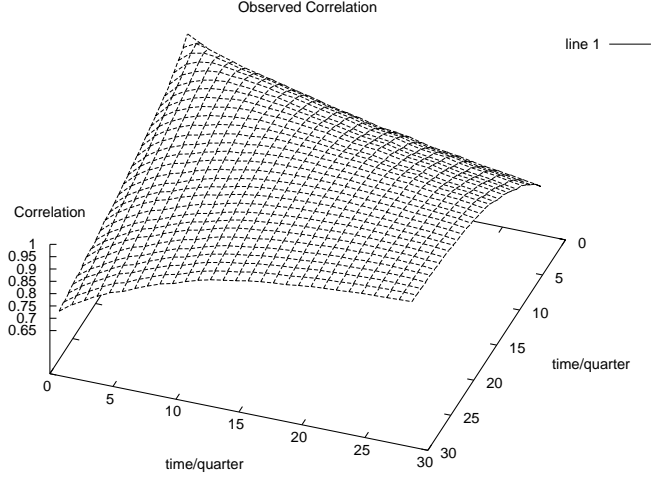, angle=-90, width=10cm}
  \caption{Empirical correlation $\label{correlator}
{\cal C}(\theta, \theta') = \frac{\langle\delta f(t, \theta) \delta
  f(t, \theta')\rangle_c }{\sqrt{\langle\delta f^2(t,
    \theta)\rangle_c}
  \sqrt{\langle\delta f^2(t, \theta')\rangle_c}}$, determined 
from the Eurodollar rates in the period 1994-1996. See \cite{data} for
more details on the data.}
  \label{fig:market_corr_one}
\end{figure}

However, as discussed in \cite{BaaqSri}, the surface of the empirical 
propagator  
given in Figure \ref{fig:market_corr_one}, is extremely smooth and shows no such kinks. This observation is in fact related to the empirical study of \cite{data} (see also \cite{data1,Risks}),
where the correlation of the (fixed maturity) forward rates daily variations $\delta f(t,\theta)\equiv  f(t+\epsilon,t+\epsilon+\theta)- 
f(t,t+\theta)$ was studied. More precisely, the eigenvectors $\Psi_q(\theta)$ and eigenvalues $\zeta_q$, $q=1,2,...$, 
of the correlation matrix ${\cal N}(\theta, \theta')$,
defined as
\begin{eqnarray}
{\cal N}(\theta, \theta')&=&< \delta f(t,\theta)\delta f(t,\theta')>_c \nonumber\\
                         &\approx &\epsilon \sigma(\theta)\sigma(\theta') D(\theta,\theta')
\nonumber
\end{eqnarray}
were determined. It was found that the eigenvectors show a structure similar to the modes of a vibrating string 
($\Psi_1$ has no nodes, $\Psi_2$ has one node, etc. - see also \cite{Rama}), 
and that the eigenvalues $\zeta_q$ behave as $(a + bq^2)^{-1}$
for small $q$ (where $a,b$ are constants), crossing over to a faster decay $\approx q^{-4}$ for larger $q$'s. 
It is clear that the square gradient action does indeed lead to plane-wave eigenvectors, with eigenvalues given
by $(a + bq^2)^{-1}$ with $b \propto \mu^{-2}$. However, the $q^{-4}$ falloff points to the existence of another 
term in the action that is a fourth power in the derivative along the future time $x$. 
This higher power of the derivative {\it stiffens} the fluctuations of the forward rates curve, in the sense that
two nearby maturities experience more correlated external shocks. The aim of the rest of this letter is to 
compare in details empirical data with the predictions of the following action
\begin{equation}
\label{quar}
S_Q[A]=-\frac{1}{2}\int_{t_0}^{\infty}dt\int_t^{\infty}
dx\left\{A^2(t,x)+\frac{1}{\mu^2}\left(\frac{\partial A(t,x)}{\partial
x}\right)^2+ \frac{1}{\lambda^4}\left(\frac{\partial^2 A(t,x)}{\partial
x^2}\right)^2\right\},
\end{equation}
that includes the new stiffness term. We will see that this theory indeed allows one to get rid of the infinite 
curvature of the propagator along the diagonal, but that a quantitative agreement with empirical data can only
be achieved if the action is written in terms of an effective `psychological' time $z$ in the maturity direction, that
is a sublinear function of the `true' maturity $x-t$. In other words, changes in the maturity direction are given by $\partial/\partial z$. 
The introduction of $z$ is similar (but not equal \cite{BaaqSri}) to the rigidity $\mu^{-2}$ and the stiffness 
$\lambda^{-2}$  not being constant along the $x$ direction, which should be expected. With this extra ingredient, we
will see that the details of the propagator surface are reproduced with surprising accuracy. 

Let us first compute the propagator of the `stiff' action above. Using eq.(\ref{lart}), we find
\begin{eqnarray}
\label{propquar}
G(\theta_+;\theta_-) = \left(\frac{\lambda^4}{\alpha_+-\alpha_-}\right)
\big [\frac{1}{\alpha_{-}}D(\theta_+;\theta_-;\sqrt{\alpha_{-}})-\frac{1}{\alpha_{+}}
D(\theta_+;\theta_-;\sqrt{\alpha_{+}})\big ]
\end{eqnarray}
with
\begin{equation}
\alpha_{\pm}=\frac{\lambda^4}{2\mu^2}\left[1\pm 
\sqrt{1-4(\frac{\mu}{\lambda})^4}~~\right]\nonumber
\end{equation}
In the limit $\lambda \rightarrow \infty$ one finds $\alpha_{+} \simeq \lambda^4/\mu^2 $ and 
$\alpha_{-}\simeq \mu^2$. Hence the propagator has the following limit
\begin{equation}
\lim_{\lambda \rightarrow \infty}G(\theta_+;\theta_-;\mu,\lambda)\rightarrow D(\theta_+;\theta_-;\mu) \nonumber
\end{equation}
and, as expected, reduces to the case of constant rigidity.

The solution for $\alpha_{\pm}$ yields three distinct cases, namely, when $\alpha_{\pm}$ is  real, complex or 
degenerate depending on whether $\mu<\sqrt{2}\lambda,  \mu>\sqrt{2}\lambda,  \mu=\sqrt{2}\lambda$ respectively. One finds
\begin {eqnarray}
G(\theta_+;\theta_-) =~~~~~~~~~~~~~~~~~~~~~~~~~~~~~~~~~~~~~~~~~~~~~~~~~~~~~~~~~~~\nonumber\\ \nonumber\\
\label{stfprop}  
\left\{ \begin{array}{ll}
\frac{\lambda}{2\sinh(2b)}\big [e^{-\lambda \theta_+\cosh(b)}\sinh\{b+\lambda\theta_+ \sinh(b)\}+
e^{-\lambda|\theta_-| \cosh(b)}\sinh\{b+\lambda|\theta_-| \sinh(b)\}\big ] & \\ \\
\frac{\lambda}{4}\big [e^{-\lambda\theta_+}\{1+\lambda \theta_+\}+ 
e^{-\lambda|\theta_-|}\{1+\lambda |\theta_-|\}\big ] & \\ \\
\frac{\lambda}{2\sin(2\phi)}\big [e^{-\lambda\theta_+ \cos(\phi)}\sin\{\phi+\lambda\theta_+ \sin(\phi)\}+
e^{-\lambda|\theta_-| \cos(\phi)}\sin\{\phi+\lambda|\theta_-| \sin(\phi)\}\big ]
 &
\end{array}
\right.    
\end {eqnarray}
In the above equation, $\alpha_{\pm}=\lambda^2e^{\pm b}$ in the first case, and $\alpha_{\pm}=\lambda^2e^{\pm i\phi}$
in the third case, and $b=\phi=0$ in the degenerate case. Expanding the propagator $G(\theta_+;\theta_-)$ 
about $\theta_{-}=0$ leads to a cancellation of the term linear in $|\theta_-|$ and gives a final result that 
is a function of $\theta_-^2$. More precisely, the curvature $r_Q$ orthogonal to the diagonal line $\theta_-=0$ 
is given, in the real case, by
\begin{equation}
\label{curq}
r_Q=\frac{\partial^2 G_b(\theta_{+};\theta_{-})}
{\partial \theta_{-}^2}|_{\theta_{-}=0} = 
-\frac{\lambda^3 \sinh(b)}{2\sinh(2b)}\big[\cosh^2(b)-\sinh(b)\big]<0.
\end{equation}
$r_Q<0$ follows from the fact that $b\ge 0$, confirming that the value of the propagator 
along $\theta_-=0$ is a maximum.
A similar result holds in the complex case. 
Note that in the limit of $\lambda \to \infty$, one can no longer carry out 
the Taylor expansion around $\theta_-=0$,  
the cancellation of the term linear in $|\theta_-|$ becomes invalid, 
and the propagator $G(\theta_+;\theta_-)$ develops the expected kink.

\begin{figure}[h]
  \centering
  \epsfig{file=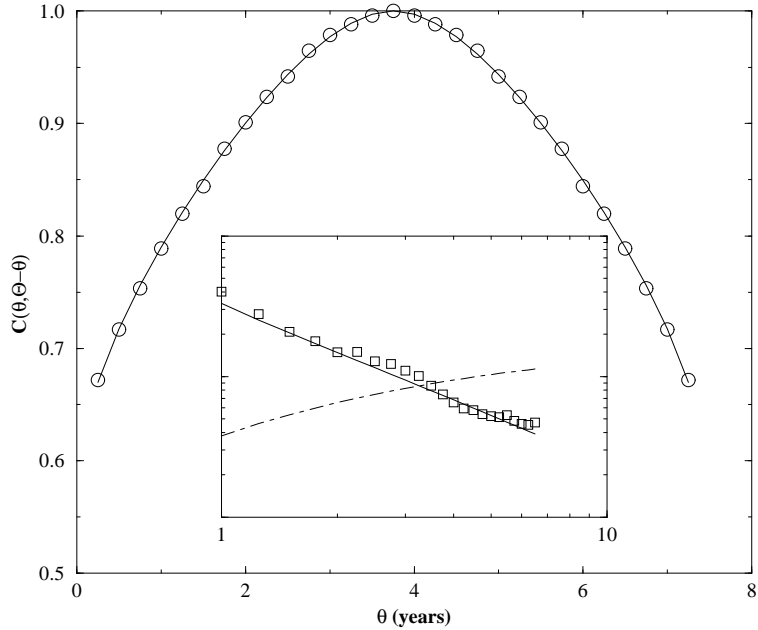, angle=-90, width=10cm}
  \caption{Circles: Empirical correlation ${\cal C}(\theta;\theta')$ along the longest stretch perpendicular to the 
  diagonal, i.e. $\theta'=\Theta-\theta$, where $\Theta$ is the maximum available maturity. The plain line 
  is the best fit with our stiff propagator model and a power-law psychological time. The inset shows a plot,
  in log-log coordinates, of the curvature $-R(\theta_+)$ as a function of $\theta_+$, 
   for a) the stiff model in real (physical) time, showing an increasing curvatue (dashed line)
  and b) the stiff model with a power law psychological time $z(\theta)$, correctly sloping downwards as to reproduce 
  the power law behaviour of the empirical curvature (squares).}
  \label{qrtcdercorr}
\end{figure}

In order to compare with empirical data, we define the normalized correlation function by
\begin{equation}
{\cal C}(\theta,\theta')=\frac{G(\theta,\theta')}{\sqrt{G(\theta,\theta)G(\theta',\theta')}}
\end{equation}
Of special interest will be the curvature of ${\cal C}(\theta,\theta')$ 
perpendicular to the diagonal, for
which an explicit expression can be obtained. 
We do not show this formula here but note that the curvature
is predicted to {\it increase} with $\theta_+$ (see Fig \ref{qrtcdercorr}, inset). 
As one moves along the diagonal to 
longer maturities, the (negative) curvature of ${\cal C}(\theta,\theta')$ increases, 
which means that 
the noise affecting nearby maturities is faster to decorrelate as a function of 
$\theta-\theta'$. This is 
contrary to intuition: since the long term future is much more uncertain, one feels 
that shocks in the far future 
are more difficult to resolve temporally than shocks in the near future. 
Therefore one expects, and indeed
empirically find, that the curvature is a decreasing function of the maturity. 
In fact, we have discoved the new
following `stylized fact': the curvature $R$ of the 
FRC correlation function along the diagonal decays as a power 
law of the maturity, $R(\theta_+) \sim \theta_+^{-\nu}$, with 
$\nu \approx 1.32$ (see Fig \ref{qrtcdercorr}, inset).

A way to reconcile the above finding with our stiff quantum field theory is to realize that the gradient 
terms in the action need not be uniform. More intuitively, in the mind of market participants, 
the perceived time between events depends on the time in the future, and a decreasing function of the maturity itself. 
The distance (in time) between - say - a 10 years maturity and a 30 years maturity is clearly much less than 
the distance between a 1 month maturity and a 10 year maturity.
A way to describe this mathematically is to replace the true, physical future time $\theta=x-t$ by a 
{\it psychological} time $z=z(\theta)$ \cite{BaaqSri,sthesis}, which is expected to grow sublinearly with $\theta$,
so that the rate of time change $z'(\theta)$ is a decreasing function of $\theta$. 
In general, one can impose some general features of function $z(\theta)$: it is monotone increasing, 
such that $\theta=\theta(z)$ is well defined, and one can impose $z(0)=0~;~z(\infty)=\infty$. 
The independent variables will now be $t,z(\theta)$ instead of $t,x$.
Our final model for the forward rate dynamics then reads:
\begin{equation}
\label{dfAz}
\frac{\partial f}{\partial t}(t,\theta)=\alpha(t,z(\theta))+\sigma(t,z(\theta))A(t,z(\theta))~~;~~\theta=x-t
\end{equation}
where $f(t,\theta)$ depends only on calendar time $\theta=x-t$. An important feature of the 
defining equation above is that both future times, namely $\theta=x-t$ and 
psychological time $z(\theta)$ occur in the theory
\footnote{The theory for psychological future time can be defined entirely 
in terms of forward rates $\tilde{f}(t,z(\theta))$. 
However, for imposing the martingale condition, it is necessary \cite{bebcup} to specify 
the relation between $\tilde{f}(t,z(\theta))$ and $f(t,\theta)$, and in effect one would recover Eq.(\ref{dfAz})}. 
The corresponding stiff action in psychological time is written as
\begin{equation}
\label{actqrtz}
  S_{z} = -\frac{1}{2} \int_{t_0}^{\infty} dt \int_{0}^{\infty} dz 
  \left(A^2 + \frac{1}{\mu^2} \left(\frac{\partial A}{\partial
        z}\right)^2+ \frac{1}{\lambda^4} \left(\frac{\partial^2 A}{\partial
        z^2}\right)^2\right)  
\end{equation}
The propagator for $S_{z}$ is $G(z,z';\mu,\lambda)$ as in eq.(\ref{propquar}) and the martingale condition 
for psychological time is given by \cite{bebcup} $
\alpha(t,z)=\sigma(t,z)\int_{z(0)}^{z}dz' G(z,z')\sigma(t,z')$.
Our final result on the normalized correlation now reads
\begin{eqnarray}
\label{corrq}
{\cal C}_{Qz}(\theta+;\theta_-)&=&\frac{g_+(z_+)+g_-(z_-)}
{\sqrt{[g_+(z_++z_-)+g_-(0)][g_+(z_+-z_-)+g_-(0)]}}~~~~~~~\\
z_\pm(\theta_+;\theta_-) &\equiv &z(\theta)\pm z(\theta') \nonumber
\end{eqnarray}
with, in the real case that will be of relevance for fitting the empirical data
\begin{eqnarray}
g_+(z)&=&e^{-\lambda z\cosh(b)}\sinh\{b+\lambda z \sinh(b)\}\nonumber\\
g_-(z)&=&e^{-\lambda|z| \cosh(b)}\sinh\{b+\lambda|z| \sinh(b)\}\nonumber\\
e^{\pm b}&=& \frac{\lambda^2}{2\mu^2}\left[1\pm 
\sqrt{1-4(\frac{\mu}{\lambda})^4}~~\right]\ \nonumber
\end{eqnarray}

It can be shown \cite{bebcup} that the curvature with psychological time reads
\begin{equation}
\label{curqz2}
\frac{\partial^2{\cal C}_{Qz}(\theta_+;0)}{\partial \theta_-^2}=[z'(\theta_+)]^2R_Q(2z(\theta_+/2)),
\end{equation}
where $R_Q$ is the curvature of the model in physical time. Since one observes a power law fall off for 
the curvature, we can make the ansatz $z(\theta)=\theta^\eta$ for fitting the data. 
Using the fact that $R_Q(2z(\theta_+/2))$ varies very slowly as a function of $\theta_+$, 
one can make the following approximation
\begin{equation}
[z'(\theta_+)]^2 \propto \frac{1}{\theta_+^{\nu}} \Rightarrow 2\eta-2 = - \nu, 
\end{equation}
leading to $\eta \approx 0.34$.
Therefore the psychological time flows, as expected, much slower than real time. The rate of change of psychological
time decreases as $\approx \theta^{-0.66}$: 
a year after ten years looks similar to two weeks after a month.  

Having used the behaviour of the curvature to fix the value of $\eta$ (and thus, up to an 
irrelevant overall scale, 
the function $z(\theta)$, we are left with only two parameters, 
$\lambda$ and $\mu$, to fit the whole correlation surface ${\cal C}(\theta,\theta')$.
For the Eurodollar data that we have used (see \cite{data} for more details), we have up to 30 different maturities 
and therefore 405 different points 
(the diagonal values are trivial). We determine $\lambda$ and $\mu$ such as 
to minimize the average square error between the empirical ${\cal C}(\theta,\theta')$ and the prediction of the
model. Defining $\tilde{\lambda}^\eta=\lambda$ and $\tilde{\mu}^\eta=\mu$, the best fit is obtained for $\tilde{\lambda} = 1.79/\mathrm{year}$ and 
$\tilde{\mu} = 0.403/\mathrm{year}$,
corresponding to $b=0.845$. The residual error is as low as $0.4\%$ per point, and the order 
of magnitude of the time scales (one year) defined by $\tilde{\lambda}$ and $\tilde{\mu}$ are very reasonable. 
The remarkable 
quality of the fit can be checked in Fig.\ref{qrtcdercorr}, along the longest stretch perpendicular to the diagonal, 
i.e. $\theta'=\Theta-\theta$, where $\Theta$ is the maximum available maturity (7.5 years). 
Even more remarkable 
is that the curvature of ${\cal C}(\theta,\theta')$ along the diagonal is very precisely reproduced by the same fit, as
shown in the inset of Fig. \ref{qrtcdercorr}. Testing the fit on the second derivative of the fitted surface is of
course much more demanding. The existence of the boundary at $x=t$, or $\theta=0$, is reflected in the $\theta_+$ term 
in the propagator; if one removes this term, and in effect assumes that the forward rates exist for all 
$-\infty \le x \le +\infty$, then the fit deteriorates with the root mean square error increasing from $0.40\%$ 
to $0.53\%$. The existence of the boundary at $x=t$ can hence be seen to have a significant effect on the correlation 
of the forward rates.

Let us summarize what we have achieved in this study. The simplest field theoretical description of
the multivariate statistics of forward rate variations over time $t$ and maturities $\theta$, 
involves a quadratic action containing a gradient squared rigidity
term \cite{Baaquie}, that captures the one dimensional string nature of the forward rate curve. However, this choice leads to a 
spurious kink (infinite 
curvature) of the normalized correlation function along the diagonal $\theta=\theta'$. Motivated by a previous 
empirical study \cite{data} where the short wavelength fluctuations of the FRC were shown to be strongly reduced 
as compared to that of an elastic string, we have considered an extended action that contains a squared Laplacian
term, which describes the bending stiffness of the FRC. In this formulation, the infinite curvature singularity 
is rounded off. In order to fit to the observed correlation functions of the forward interest rates, however, 
one has to add as an extra ingredient that the rigidity/stiffness constants are in fact not constant along the 
maturity axis but increase with maturity. An intuitive and parsimonious way to describe this effect is to postulate 
that markets participants, who generate the random evolution of the FRC, do not perceive future time in a 
uniform manner. Rather, time intervals in the long term future are shrunk. The introduction of a `psychological time'
$z(\theta)$, found to be a power law of the true time, allows one to provide an excellent fit of empirical data, 
and in particular to reproduce accurately a new stylized fact: the curvature of the forward rate correlation
perpendicular to the diagonal decays as a power-law of the maturity. We believe that our quantum field formulation, 
including the new stiffness term and coupled with an appropriate calibration of the term structure of the 
volatility $\sigma(\theta)$ (see \cite{data,data1}), accounts extremely well for the phenomenology of interest 
rate dynamics. It is also mathematically tractable, and should allow one to compute in closed forms derivative prices and optimal
hedging strategies. It would be interesting to generalize the above model to account for non Gaussian effects,
that are important in many cases \cite{Risks}. This would amount to considering non quadratic terms in $A$ in the action. 
We leave this extension for future work.

\end{document}